\documentclass[12pt,preprint]{aastex}         
\usepackage{graphicx}
\usepackage{epstopdf}
\usepackage{amsmath}
\def\be{\begin{equation}}
\def\ee{\end{equation}}

\def\lsim{\lower 2pt \hbox{$\, \buildrel {\scriptstyle <}\over
         {\scriptstyle \sim}\,$}}

\begin{document}
\newcommand{\figureout}[3]{\psfig{figure=#1,width=5.5in,angle=#2} 
   \figcaption{#3} }

\title{Multiwavelength Polarization of Rotation-Powered Pulsars}

\author{Alice K. Harding\altaffilmark{1} and Constantinos Kalapotharakos\altaffilmark{1}$^{,}$\altaffilmark{2}} 
  
\altaffiltext{1}{Astrophysics Science Division,      
NASA Goddard Space Flight Center, Greenbelt, MD 20771}
\altaffiltext{2}{NASA Postdoctoral Fellow, USRA}
 

\begin{abstract}
Polarization measurements provide strong constraints on models for emission from rotation-powered pulsars.   We present multiwavelength polarization predictions showing  that measurements over a range of  frequencies can be particularly important for constraining the emission location, radiation mechanisms and system geometry.  The results assume a generic model for emission from the outer magnetosphere and current sheet in which optical to hard X-ray emission is produced by synchrotron radiation from electron-positron pairs and $\gamma$-ray emission is produced by curvature radiation or synchrotron radiation from accelerating primary electrons.   The magnetic field structure of a force-free magnetosphere is assumed and the phase-resolved and phase-averaged polarization is calculated in the frame of an inertial observer.  We find that large position angle swings and deep depolarization dips occur during the light curve peaks in all energy bands.  For synchrotron emission, the polarization characteristics are strongly dependent on photon emission radius with larger, nearly $180^\circ$, position angle swings for emission outside the light cylinder as the line-of-sight crosses the current sheet.  The phase-averaged polarization degree for synchrotron radiation is less that 10\% and around 20\% for emission starting inside and outside the light cylinder respectively, while the polarization degree for curvature radiation is much larger, up to 40\% - 60\%.  Observing a sharp increase in polarization degree and a change in position angle at the transition between X-ray and $\gamma$-ray spectral components would indicate that curvature radiation is the $\gamma$-ray emission mechanism.  
\end{abstract} 


\pagebreak
  
\section{Introduction}

High-energy emission from rotation-powered pulsars has been detected and studied for several decades, most recently by the Fermi Gamma-Ray Space Telescope and a number of X-ray telescopes.  It is now clear that most of the pulsed non-thermal X-ray and $\gamma$-ray emission comes from the outer magnetosphere of the pulsars.  Various models that assume different emission locations and mechanisms are capable of producing the high-energy light curves.  In outer gap (OG, Romani 1996, Hirotani 2006) and slot gap (SG, Harding et al. 2008) models, the radiation is produced inside the light cylinder, in narrow accelerating gaps along the last open field lines starting either at the neutron star surface for the SG or above the null charge surface for the OG.  OG and SG models both find that non-thermal X-rays are synchrotron emission from electron-positron pairs while $\gamma$-ray emission comes from curvature radiation of primary accelerated particles.  Other models invoke emission outside the light cylinder near the current sheet, but there is disagreement over whether the high-energy $\gamma$ rays come from synchrotron radiation (Petri \& Kirk 2005, Cerutti et al. 2016), curvature radiation (Kalapotharakos et al. 2014) or inverse Compton emission (Lyutikov 2013).

Polarization signatures on the other hand can distinguish between different emission mechanisms and emission locations in the pulsar magnetosphere.  Since the pulsed radiation is emitted along the direction of relativistic particle motion  which is coupled to the magnetic field direction, and the electric vector, parallel to the particle acceleration, is different for curvature and synchrotron radiation, the polarization can be a very sensitive diagnostic.  Dyks et al. (2004) made phase-resolved polarization predictions for geometric renditions of the OG and two-pole caustic (TPC, Dyks \& Rudak 2003, similar to the SG) models, showing that fast swings of the position angle (PA) and dips in the polarization fraction (P) during the pulse peaks are expected.  In these models, the pulse peaks are caustics in the emission pattern caused by relativistic effects of aberration and light travel time that compress the arrival times of photons emitted along trailing field lines into a single phase.  The addition of radiation from a large range of altitudes, and thus field directions, produces the fast PA swings and depolarization at the peaks.  They found that polarization of  TPC model profiles with two peaks showed large PA swings and P dips during both peaks while the OG model showed a much smaller PA swing at the first peak and depolarization both at and between the peaks.  Takata \& Chang (2007) modeled multiwavelength polarization signatures in the OG model, and found that extending the emission below the null surface produces a large PA swing at the first peak and confirmed the depolarization throughout the bridge region between the peaks for synchrotron emission.   Petri \& Kirk (2005) and Petri (2013) explored polarization of synchrotron emission in the striped wind at the current sheet outside the light cylinder where the peaks are produced when the observer line-of-sight crosses the current sheet of an oblique rotator.  Since the magnetic field drops to zero and changes polarity at the current sheet, the PA is predicted to swing by up to 180 degrees and the P drops at the peaks.  Contopoulos \& Kalapotharakos (2010) computed the phase-resolved polarization of curvature radiation from the current sheet of a force-free magnetosphere and predicted a similar behavior.
Cerutti et al. (2016), using results of a particle-in-cell code for the global pulsar magnetosphere to model synchrotron radiation from the current sheet, find similar swings of PA at the pulse peaks, although the depolarization dips are less pronounced. 

Polarization measurements at optical energies and above exist for only a few rotation-powered pulsars.  The Crab pulsar is the best studied, with phase-resolved polarimetry at optical wavelengths (Slowikowska et al. 2009), showing fast swings of PA and depolarization at both pulse peaks.  Soft $\gamma$-ray (Forot et al. 2008) polarization measurements in several phase bins with INTEGRAL/IBIS show lower polarization fraction in the peaks than in the bridge and off-peak regions, which is roughly consistent with the optical behavior.  Polarization measurement of the Crab at X-ray energies by Weisskopf et al. (1978) is limited to the total nebula plus pulsar emission.  For other pulsars, including Vela (Moran et al. 2014), PSR B0540-69 (Mignani et al. 2010), PSR B0656+14 (Kern et al. 2003) and PSR B1509-58 (Wagner \& Seifert 2000), only optical/UV polarization measurement have been made.  The phase-averaged polarization degree of the Crab, Vela, PSR B0540-69 and PSR B1509-58 are all in the range 10\% - 16\% and the position angles (measured for all but PSR B1509-58) are aligned with the pulse spin axis and/or the proper motion direction.  The phase-resolved PA of PSR B0656+14 shows a rotating vector model pattern typical of radio emission, and a very high polarization degree, indicating that the optical emission may come from near the polar cap rather than from the outer magnetosphere.

In this paper, we model the phase-averaged and phase-resolved polarization of pulsar emission over a broad energy range from the outer magnetosphere and current sheet using a force-free magnetic field geometry.  Our emission model includes synchrotron radiation at optical to hard X-ray energies and curvature or synchrotron radiation at $\gamma$-ray energies, taking into account the intrinsic degree of polarization and position angle for each radiation process as well as the emission geometry and relativistic effects that influence the pulsations.  We compare our results to optical and hard X-ray measurement of the Crab pulsar and argue that measurement of the polarization in the $\gamma$-ray band could distinguish between a synchrotron or curvature origin for this emission.  In \S \ref{sec:des} we describe the radiation model and polarization calculation.  In \S \ref{sec:results} we present our results on phase-averaged and phase-resolved polarization, and in \S \ref{sec:diss} we compare our results to existing observations as well as discuss our predictions for future high-energy polarimeters.

\section{Description of Models}  \label{sec:des}

\subsection{Radiation Modeling}  \label{sec:mod}

We use a simplified version of the radiation model presented by Harding \& Kalapotharakos (2015, HK15), who simulated synchrotron radiation (SR), curvature radiation (CR) and synchrotron self-Compton emission (SSC) from a population of primary electrons and electron-positron pairs moving away from the neutron star polar cap.  The particles originate at points on the neutron star surface that fill concentric rings near the edge of the polar cap.   The primary electrons are injected with a low Lorentz factor ($\gamma$ = 2000) and accelerated by a constant electric field component parallel to the magnetic field, $E_\parallel$ from the neutron star surface up to maximum radius $r_{\rm max} = 2.0\,R_{\rm LC}$, where $R_{\rm LC} = c/\Omega$ is the light cylinder radius, and their energies are limited by CR reaction, while the pairs are injected with the energy distribution of the polar cap cascade spectrum computed for the Crab pulsar (shown in Figure 1 of HK15) and assumed not to be accelerated.  Since the details of this model are described in HK15 we will note here only the modifications we have made for the polarization calculation of this paper.  First, we simulate here only SR from a spectrum of pairs and CR or SR from primary electrons, ignoring SSC since it appears at very high energies $> 1$ GeV.  Second, the treatment of SR is greatly simplified.  In HK15, the particles gained pitch angles through cyclotron resonant absorption of radio photons and suffered SR losses, which is time-consuming to compute.  Here, we simply assume that the pairs, with a range of Lorentz factors $\gamma$,  have zero momentum perpendicular to the magnetic field, $p_\perp$, from the start of their trajectories up to a radius $r_{\rm min}$ and maintain a constant $p_\perp$ above a radius $r_{\rm min}$ to maximum radius $r_{\rm max}$.  We explore different constant pitch angles $\psi =  p_\perp/p$.  In this way, we can explore how the polarization characteristics change for radiation at different altitudes.  Third, we modulate the particle flux in each ring by a Gaussian of width $\Delta r = 0.05$ centered at $r_0 = 0.9$ in units of polar cap radius, which is the same for primaries and pairs,  while HK15 assumed uniform particle flux between inner and outer rings at $r_{\rm in}$ and $r_{\rm out}$ respectively that had different values for primaries and pairs.  Use of a Gaussian profile produces smoother light curves and phase-resolved polarization.   As in HK15, the photon flux from the injected test particles are normalized to the Goldreich-Julian flux, $\dot n_{\rm GJ} = n_{\rm GJ}c A_{\rm inj}$, where $n_{\rm GJ} = B_0 \Omega/2\pi ec$ is the Goldreich-Julian number density at the neutron star surface and $A_{\rm inj}$ is the area on the polar cap over which the injection occurs.

Computation of particle trajectories and radiation is done in the inertial observers frame (IOF) since the force-free magnetic field structure has been determined through numerical calculations  in the IOF (c.f. Kalapotharakos \& Contopoulos 2009, Kalapotharakos et al. 2012).   As described in HK15, we extend the numerical solution to a neutron star surface of $0.01 R_{\rm LC}$ by joining it to the analytic Deutsch (1955) retarded vacuum dipole solution below $0.2 R_{\rm LC}$, using a ramp function between 0.2 and $0.4 R_{\rm LC}$.  At each location above $0.2 R_{\rm LC}$ we interpolate the magnetic field components from stored numerical table values.  We assume all photons are emitted along the direction of the particle velocity and the Stokes Parameters for the emission (see \S \ref{sec:pol}) at steps along the particles trajectories are accumulated in sky maps for 24 energy intervals.  

All the parameters for these calculations are set for the Crab pulsar, with period $P_{\rm rot} = 0.033$ s and period derivative $\dot P_{\rm rot} = 4.22 \times 10^{-13}\,\rm s\,s^{-1}$, and the pair multiplicity is set to $M_+ = 10^5$.  The results are most applicable to the Crab and Crab-like pulsars such as PSR B0540-69 and PSR B1509-58 that show higher levels of optical and non-thermal X-ray emission and thus are more likely to have detectable phase-averaged and phase-resolved polarization measurements.  

\subsection{Calculation of Polarization} \label{sec:pol}

To determine the polarization of the emitted radiation, we calculate the Stokes Parameters at each step along the particle trajectories for each spectral interval,
\begin{align}
& I(\omega) = N_{CR} (\omega) + N_{SR} (\omega) \nonumber
\\
& Q(\omega) = N_{CR} (\omega) \cos(2\chi_{_{CR}}) P_{CR}(\omega) + N_{SR} (\omega) \cos(2\chi_{_{SR}}) P_{SR}(\omega)  \nonumber
\\
& U(\omega) = N_{CR} (\omega) \sin(2\chi_{_{CR}}) P_{CR}(\omega)+ N_{SR} (\omega) \sin(2\chi_{_{SR}}) P_{SR}(\omega)
\end{align}
where $N_{CR}(\omega)$ and $N_{SR}(\omega)$ are the spectral flux, $\chi_{_{CR}}$ and $\chi_{_{SR}}$ are the position angles, and $P_{CR}(\omega)$ and $P_{SR}(\omega)$ are the intrinsic polarization degree of CR and SR at photon energy $\omega$.  To compute the degree of intrinsic polarization of both SR and CR, we use the expression (Westfold 1959)
\be
P(\omega) = {K_{2/3}(\omega/\omega_c)\over \int_{\omega/\omega_c}^\infty K_{5/3}(x)dx}
\ee
where $K_{2/3}(x)$ and $K_{5/3}(x)$ are modified Bessel functions and $\omega_c$ are the critical energies for CR
\be
\omega_{CR} = {3\over 2} {c\,\gamma^3\over \rho_c}
\ee
and for SR
\be 
\omega_{SR} = {3\over 2} \gamma^2\,B'\,\sin\psi
\ee
where $\gamma$ is the particle Lorentz factor, $\rho_c$ the particle trajectory radius of curvature, $B'$ the local magnetic field strength in units of the critical field, $B_{\rm cr} = 4.4 \times 10^{13}$ G, and $\psi$ is the particle pitch angle.  Asymptotic values of $P(\omega)$ are 0.5 for $\omega \ll \omega_c$ and 1.0 for $\omega \gg \omega_c$.  At $\omega = \omega_c$, $P \simeq 0.75$ (Westfold 1959).

The direction of polarization for each process is determined by the orientation of the photon electric vector which is parallel to the direction of particle acceleration (Blaskiewicz et al. 1991, Hibschman \& Arons 2001).  The direction of the particle acceleration in the IOF is the derivative of the velocity (Takata et al. 2007),
\be \label{eq:beta}
\boldsymbol{\beta} = \boldsymbol{\beta_d} + f(\cos\psi {\bf \hat b} + \sin\psi\cos\varphi_g  {\bf \hat n}+ \sin\psi\sin\varphi_g  {\bf \hat k})
\ee
where $\boldsymbol{\beta_d}$ is the particle drift velocity component
\be
\boldsymbol{\beta_d} = \frac{\mathbf{E}\times\mathbf{B}}{B^2},
\ee
$\psi$ is the particle pitch angle, and $\varphi_g$ is the gyro phase. By requiring $\beta = 1$ and outward motion,  the scalar quantity $f$ can be determined at each point along the trajectory.  The unit vector  ${\bf \hat b} = {\bf B}/B$ is the direction of the local magnetic field.  The unit vectors ${\bf \hat n}$ and ${\bf \hat k}$ are orthogonal to ${\bf \hat b}$, and are oriented such that ${\bf \hat n} = ({\bf \hat b} \cdot \nabla){\bf \hat b}/|({\bf \hat b} \cdot \nabla){\bf \hat b}|$ is in the plane of the field line curvature and  ${\bf \hat k} = {\bf \hat n \times \bf \hat b}$.  In the case of CR, $\psi = 0$ and we determine the particle acceleration direction, and thus the electric vector 
${\bf \hat e_{\rm CR}}$, through a numerical derivative of the particle velocity which is the change in direction of the trajectory at two successive positions.  
In the case of SR, one would need to resolve the particle gyro motion with $\varphi_g = \varphi_0 + \Omega_B t$ in order to numerically compute the velocity derivative, where $\Omega_B = eB/\gamma mc$ is the gyro frequency.  Since it is impractical to resolve the gyro motion in the high pulsar magnetic field, we can use the analytic derivative of the particle velocity in Eq. (\ref{eq:beta}) (Takata et al. 2007),
\be
{\bf a_{\rm SR}} = f \Omega_B \sin\psi (\cos\varphi_g {\bf \hat k} - \sin\varphi_g  {\bf \hat n})
\ee

\noindent which is a very good approximation if $\Omega_B \gg \Omega$.  The SR electric vector direction is then ${\bf \hat e_{SR} =  a_{\rm SR}/|a_{\rm SR}|}$.  In the particle gyro motion, the phase $\varphi_g$ rotates around the magnetic field direction, changing orientation over $2\pi$ each gyro period.  This affects both the photon emission direction which is along the particle velocity (Eq. \ref{eq:beta}), which broadens the caustics in the sky map, and the direction of the radiation electric vector causing depolarization of the SR.  To capture this in our calculation, we average the emission over a range of $\varphi_g$ values, evenly spaced between 0 and $2\pi$, at each point in the trajectory.

We define the PA as the angle, in the counterclockwise direction, between the electric vector ${\bf \hat e'}$ and the projected pulsar spin axis ${\bf \Omega'}$ on the plane of the sky:
\begin{align}
&{\bf  \hat e'_{CR} = \hat e_{CR} - (\hat \eta \cdot \hat e_{CR})\,\hat \eta} \\ \nonumber
&{\bf \hat e'_{SR} = \hat e_{SR} - (\hat \eta \cdot \hat e_{SR})\,\hat \eta} \\ \nonumber
&{\bf \Omega' = \Omega  - (\hat \eta \cdot \Omega)\,\hat \eta}
\end{align}
where $\hat \eta$ is the direction of the photon emission, assumed to be the particle velocity direction.  The PA for CR and SR are then
\be
\label{eqn:psi}
\chi (\omega) = {\rm atan}\left ({\hat e'_y\over \hat e'_x}\right)
\ee
Stokes Parameters $I_{i,j,k}$, $U_{i,j,k}$ and $V_{i,j,k}$ from all particle trajectories are accumulated for each energy $\omega_i$, observer angle $\zeta_j$ and phase $\phi_k$ with respect to ${\bf \Omega}$ to form energy dependent sky maps $(\omega,\zeta,\phi)$.  We can then make sky maps of flux, 
\be
\label{eqn:F}
F(\omega,\zeta,\phi) =I_{i,j,k},
\ee
polarization fraction
\be
\label{eqn:P}
P(\omega,\zeta,\phi) = {[Q^2_{i,j,k} + U^2_{i,j,k}]^{1/2}\over I_{i,j,k}},
\ee
and position angle
\be
\label{eqn:PA}
PA(\omega,\zeta,\phi) = 0.5\, {\rm atan}\left({U_{i,j,k}\over Q_{i,j,k}}\right).
\ee
For phase-resolved polarization, cuts across the sky maps at constant $\zeta$ for a given energy range, $\omega_n$ to $\omega_m$ give the flux, polarization fraction and position angle as a function of phase, 
\begin{align}
\label{eqn:prpol}
& F_{ \zeta}(\phi)= \sum_{i=n}^m F(\omega_i,\zeta,\phi) \\ \nonumber
& P_{\zeta}(\phi)= \sum_{i=n}^m P(\omega_i,\zeta,\phi) \\ \nonumber
& PA_{ \zeta}(\phi)= \sum_{i=n}^m PA(\omega_i,\zeta,\phi)
\end{align}
We also compute the phase-averaged spectral flux, 
\be 
\label{eqn:Fw}
\langle F_{ \zeta} (\omega) \rangle = \sum_k F(\omega,\zeta,\phi_k) \Delta \phi/ 2\pi
\ee
polarization fraction
\be
\label{eqn:Pw}
\langle P_{\zeta} (\omega) \rangle = {\sum_k P(\omega,\zeta,\phi_k)\,F(\omega,\zeta,\phi_k)\over \sum_k F(\omega,\zeta,\phi_k)}
\ee
and position angle
\be
\label{eqn:PAw}
\langle PA_{\zeta} (\omega) \rangle = {\sum_k PA(\omega,\zeta,\phi_k)\,F(\omega,\zeta,\phi_k)\over \sum_k F(\omega,\zeta,\phi_k)}
\ee
where the $\langle P_{\zeta} \rangle$ and $\langle PA_{\zeta} \rangle$ have been weighted with the flux at each phase.

\section{Results}  \label{sec:results}

\subsection{Phase-Resolved Polarization} \label{sec:PRpol}

\begin{figure} 
\hskip -2.0cm
\includegraphics[width=200mm]{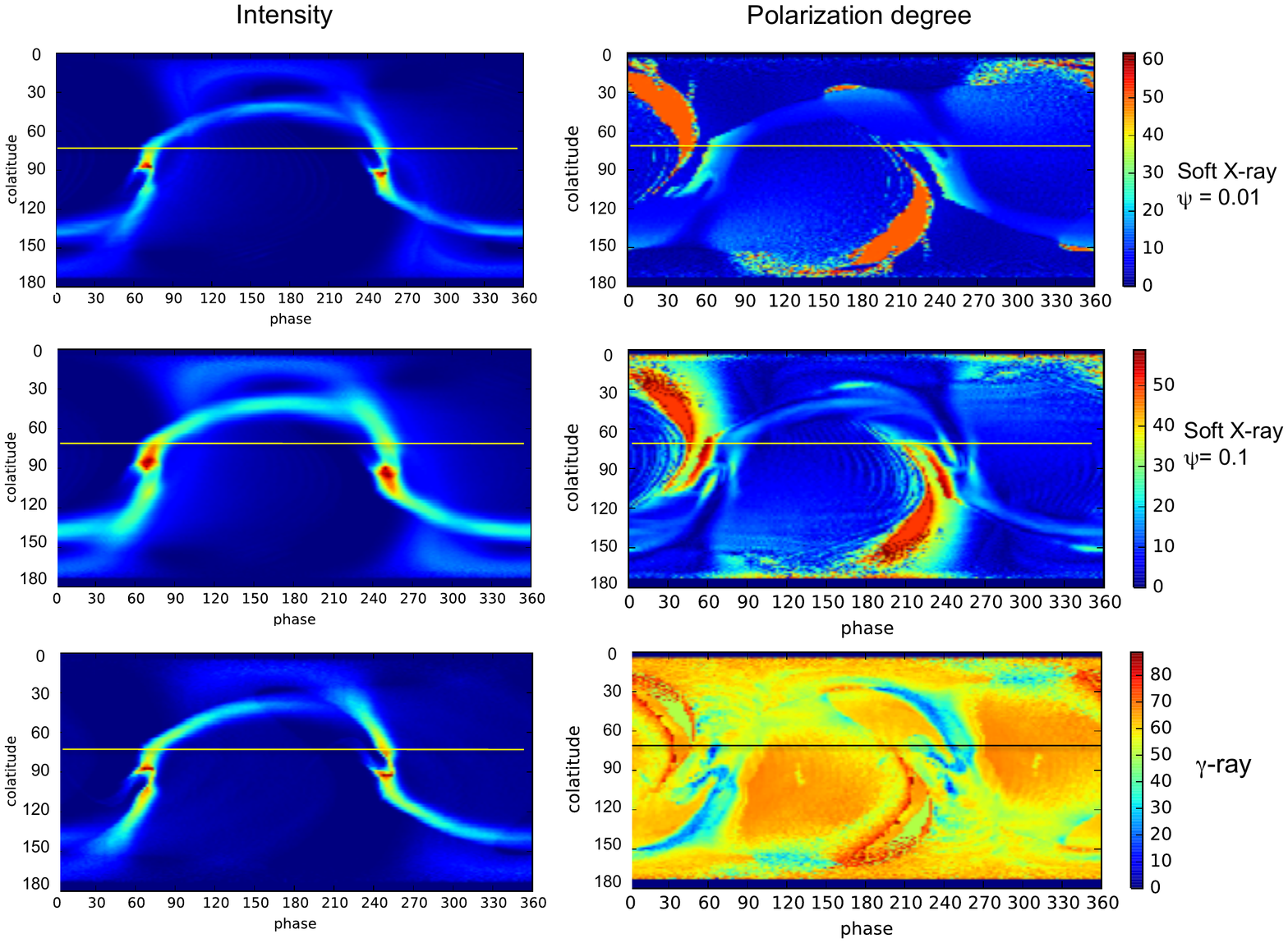}
\caption{Skymaps for $\alpha = 60^\circ$ and emission radius range $r = 0.7 - 1.3\,R_{\rm LC}$, showing emission intensity (left panels) and polarization percent (right panels) as a function of rotation phase and observer angle  (both in degrees) relative to the pulsar spin axis for soft X-ray and $\gamma$-ray energy bands.  Results for two different particle pitch angles $\psi$ are shown for the SR in the soft X-ray range.  The horizontal lines are lines of constant observer angle, $\zeta = 70^\circ$. The different color scales on the right are the corresponding percent polarization.}   
\label{fig:skymap}
\end{figure}

We present the phase-resolved polarization results for four different energy ranges, in optical ($1 - 10$ eV), soft X-ray ($2-10$ keV), hard X-ray ($0.1 - 10$ MeV) and $\gamma$-ray ($0.1 - 100$ GeV).
In Figure \ref{fig:skymap}, we show sky maps of the intensity and polarization degree for the soft X-ray and $\gamma$-ray energy bands for a pulsar inclination angle of $\alpha = 60^\circ$ and emission radius range $r = 0.7 - 1.3\,R_{\rm LC}$, derived using Eqs \ref{eqn:F} and \ref{eqn:P}.  In the soft X-ray band, the radiation is SR from pairs, shown for two different pitch angles ($\psi = 0.01$ and $0.1$), while in the $\gamma$-ray band the radiation is primarily CR from primary electrons.  The intensity pattern (left panels) shows bright caustics at locations where photons from a range of radii arrive at an observer at the same phase.  The caustic formation is somewhat different at radii inside and outside the light cylinder (LC).  Inside the LC, the particle trajectories in a force-free magnetosphere are parallel to curved, mostly poloidal magnetic field lines in the corotating frame.   In the IOF, aberration and time-of-flight produce phase delays that cancel the phase differences of emission at different radii along the trailing field lines (Morini 1983, Dyks et al. 2004), causing the emission to bunch along a preferred direction and phase toward the observer.  Outside the LC, the poloidal field lines straighten and the toroidal field becomes dominant.  The particle trajectories (defined always in the IOF) become radial and when emission occurs near the current sheet, as it does when particle trajectories lie near the last open field line or separatrix, the caustics trace the current sheet as it moves up and down in latitude with phase (Bai \& Spitkovsky 2010).   At $r = 0.7 - 1.3 \,R_{\rm LC}$, the caustics are formed partly inside and partly outside the LC.   At each observer angle $\zeta$, peaks in the light curve arise when the light-of-sight (horizontal lines) cuts once or twice across the caustic.  The range of gyro phase of SR produces different particle trajectories that broaden the soft X-ray caustics to a greater degree for larger pitch angle (Takata et al. 2007).  As is apparent, the light curve peaks are in phase for different energy ranges.  The right panels show sky maps of polarization degree for the soft X-ray and $\gamma$-ray energy ranges and display patterns of depolarization.  For emission inside the LC, the dip in polarization is due to superposition of different field directions from a large range of radii, whereas for emission outside the LC near the current sheet, the depolarization comes from superposition of emission from field lines of opposite polarity on either side of the sheet.  The polarization dips should therefore come nearly in phase with the light curve peaks.   For SR in soft X-rays, there are also large regions of depolarization following the peaks due to the range of gyro phases for which the photon electric vector points in different directions.

\begin{figure} 
\vskip -5.0cm
\hskip -2.0cm
\includegraphics[width=200mm]{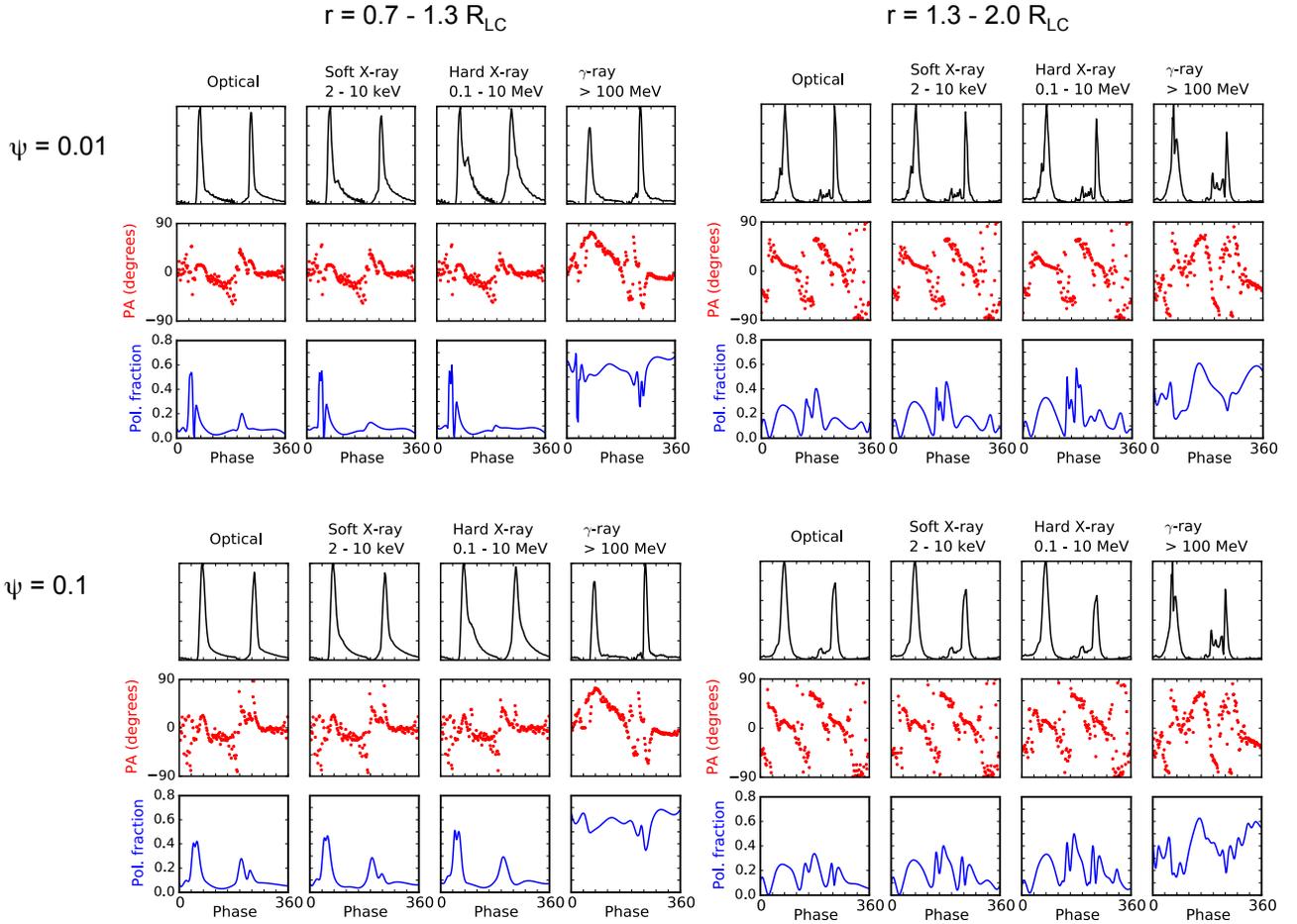}
\vskip -1.0cm
\caption{Light curves (top panels), position angle (middle panels) and polarization fraction (bottom panels) as a function of rotation phase (in degrees), for $\alpha = 60^\circ$, $\zeta = 70^\circ$, and four different energy bands - optical ($1 - 10$ eV), soft X-ray ($2-10$ keV), hard X-ray (0.1 - 10 MeV) and $\gamma$-ray ($0.1 - 100$ GeV).   Results are shown for two different emission radius ranges  $r  = 0.7 - 1.3\,R_{\rm LC}$ and $r = 1.3 - 2.0\,R_{\rm LC}$ and two particle SR pitch angles, $\psi$.  Note that the position angle swings in opposite directions for $\zeta$ and $180^\circ - \zeta$.}   
\label{fig:pr60}
\end{figure}

Figure \ref{fig:pr60} shows the light curves, PA and polarization fraction for $\alpha = 60^\circ$ and $\zeta = 70^\circ$ for different  energy bands, emission radii, $r = 0.7 - 1.3\,R_{\rm LC}$ and $r = 1.3 - 2.0\,R_{\rm LC}$ and two different pitch angles.  The light curve for this geometry shows two Crab-like peaks at the same phases at all energy bands.  The polarization of SR in the optical, soft X-ray and hard X-ray bands are nearly the same, with swings of PA in the same direction in both peaks and depolarization at each peak.  Due to the range of different gyro phases, the depolarization is quite strong with large regions of phase where the polarization is below 10\%.  The polarization of CR in the $\gamma$-ray band is quite different, with swings of PA at the peaks but much higher levels of polarization greater than 40\% even at the peaks.  In our calculation, the projection of the pulsar rotation axis on the plane of the sky is $0^\circ$ (see Eq \ref{eqn:psi}). 
For $r = 1.3 - 2.0\,R_{\rm LC}$, the PA swings are larger and the polarization dips in the $\gamma$-ray band are deeper and wider, as the position angle changes by nearly $180^\circ$ due to the change in field line polarity as the line of sight crosses the current sheet.  One interesting feature is that the PA in the bridge region for optical and X-ray bands is $90^\circ$ different from the $\gamma$-ray bridge PA for the $r = 0.7 - 1.3\,R_{\rm LC}$ case but similar to the $\gamma$-ray bridge PA for $r = 1.3 - 2.0\,R_{\rm LC}$.  This is due to the change in direction of the magnetic field as it transitions from mostly poloidal inside the LC to mostly toroidal outside the LC, so that the SR electric vector flips by nearly $90^\circ$.  On the other hand, the particle acceleration vector stays at the same orientation, so that the CR electric vector does not change from inside to outside the LC.
The PA swings for SR are also larger for the higher pitch angle case and the polarization fraction is somewhat higher between the peaks for emission at smaller radii.  Note that there is a symmetry in the PA such that the curves for $\zeta$ and $180^\circ - \zeta$ are flipped vertically about $0^\circ$.  That means that the PA swings in opposite directions for $110^\circ$ and $70^\circ$.

\begin{figure} 
\hskip -2.0cm
\includegraphics[width=200mm]{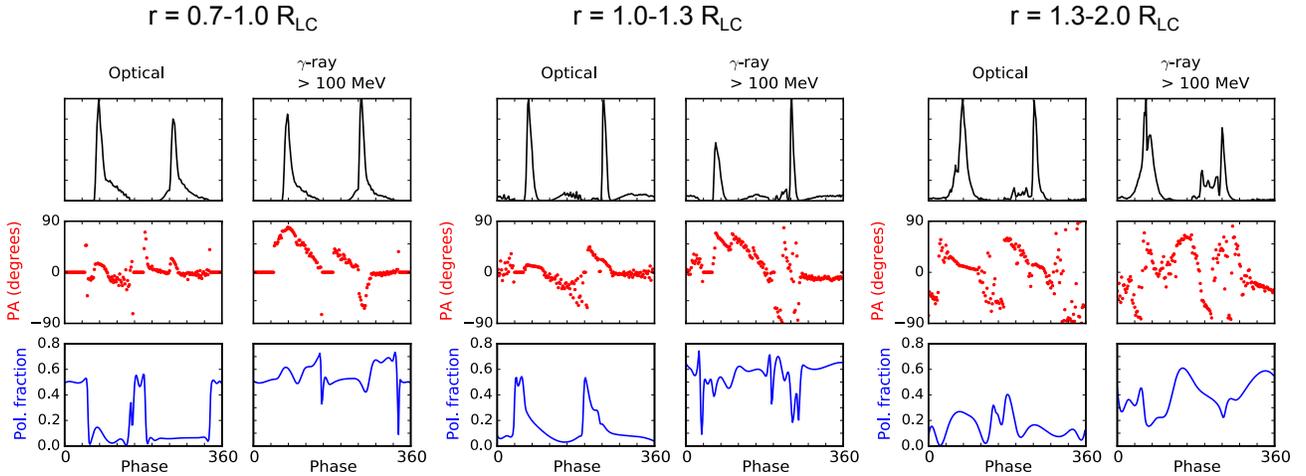}
\caption{Light curves (top panels), position angle (middle panels) and polarization fraction (bottom panels) as a function of rotation phase (in degrees), for $\alpha = 60^\circ$, $\zeta = 70^\circ$, and two different energy bands - optical ($1 - 10$ eV) and $\gamma$-ray ($0.1 - 100$ GeV).   Results are shown for three different emission radius ranges  $r  = 0.7 - 1.0\,R_{\rm LC}$, $r  = 1.0 - 1.3\,R_{\rm LC}$ and $r = 1.3 - 2.0\,R_{\rm LC}$ and SR pitch angle, $\psi = 0.01$.}   
\label{fig:pr60r}
\end{figure}

Since the polarization characteristics seem to be most sensitive to emission radius, in Figure \ref{fig:pr60r} we show the light curves, PA and polarization fraction for smaller emission ranges inside, $r = 0.7 - 1.0\,R_{\rm LC}$, and just outside the LC, $r = 1.0 - 1.3\,R_{\rm LC}$, as well as further outside the LC, $r = 1.3 - 2.0\,R_{\rm LC}$, for pitch angle $\psi = 0.01$.  Since as noted earlier there are minimal variations in the SR polarization between optical and hard X-ray bands, we show only the optical and $\gamma$-ray bands here.  There are indeed dramatic differences in polarization behavior with radius.  In general, the PA swings become larger and there is more depolarization as emission radius increases.  For emission inside the LC, the PA swings for 
SR are smallest and the polarization degree is high (50\%) outside the peaks with sudden dips to around 5\%  on either side of the peaks.  For emission just outside the LC, the regions of depolarization are wider and the polarization fraction is lower between the peaks.  For emission well outside the LC, where the current sheet is well formed, the PA swings and depolarization are maximum.

\begin{figure} 
\hskip -1.5cm
\includegraphics[width=190mm]{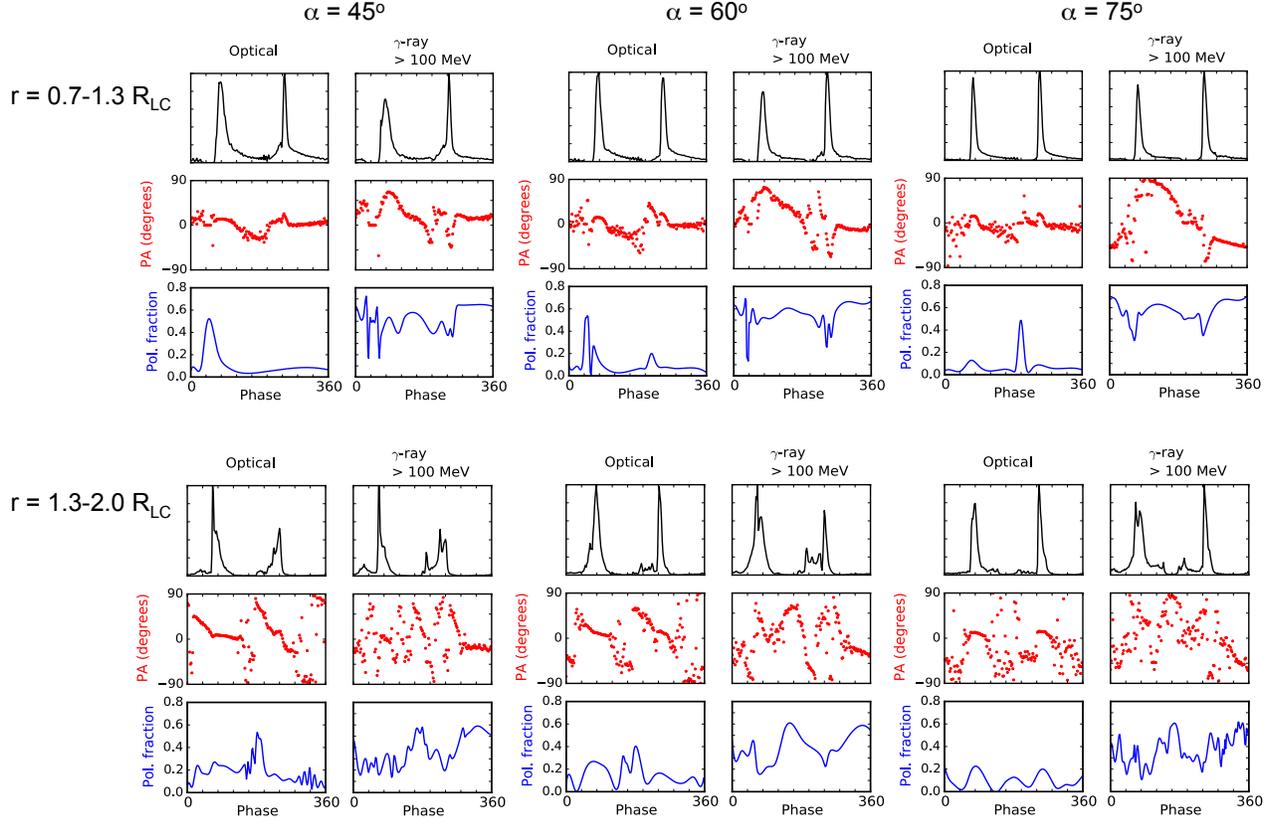}
\caption{Light curves (top panels), position angle (middle panels) and polarization fraction (bottom panels) as a function of rotation phase (in degrees), for inclination angles $\alpha = 45^\circ, 60^\circ$ and $75^\circ$, viewing angle $\zeta = 70^\circ$, and two different energy bands - optical ($1 - 10$ eV) and $\gamma$-ray ($0.1 - 100$ GeV).   Results are shown for two different emission radius ranges  $r  = 0.7 - 1.3\,R_{\rm LC}$ and $r = 1.3 - 2.0\,R_{\rm LC}$ and particle SR pitch angle, $\psi = 0.01$.}   
\label{fig:pr4575}
\end{figure}

To explore the polarization dependence on inclination angle, Figure \ref{fig:pr4575} shows the light curves, PA and polarization fraction at $\zeta = 70^\circ$ for the optical and $\gamma$-ray energy bands for $\alpha = 45^\circ, 60^\circ$ and $75^\circ$, for emission radii, $r = 0.7 - 1.3\,R_{\rm LC}$ and $r = 1.3 - 2.0\,R_{\rm LC}$.  The polarization behavior is more strongly dependent on emission radius than on inclination angle, with the swings in polarization being larger for the larger emission radii at all three inclination angles.  One noticeable trend is that at $r = 0.7 - 1.3\,R_{\rm LC}$ the PA swings for CR in the $\gamma$-ray band are larger with increasing inclination angle.

\subsection{Phase-Averaged Polarization}  \label{sec:PApol}

\begin{figure} 
\hskip -2.0cm
\includegraphics[width=200mm]{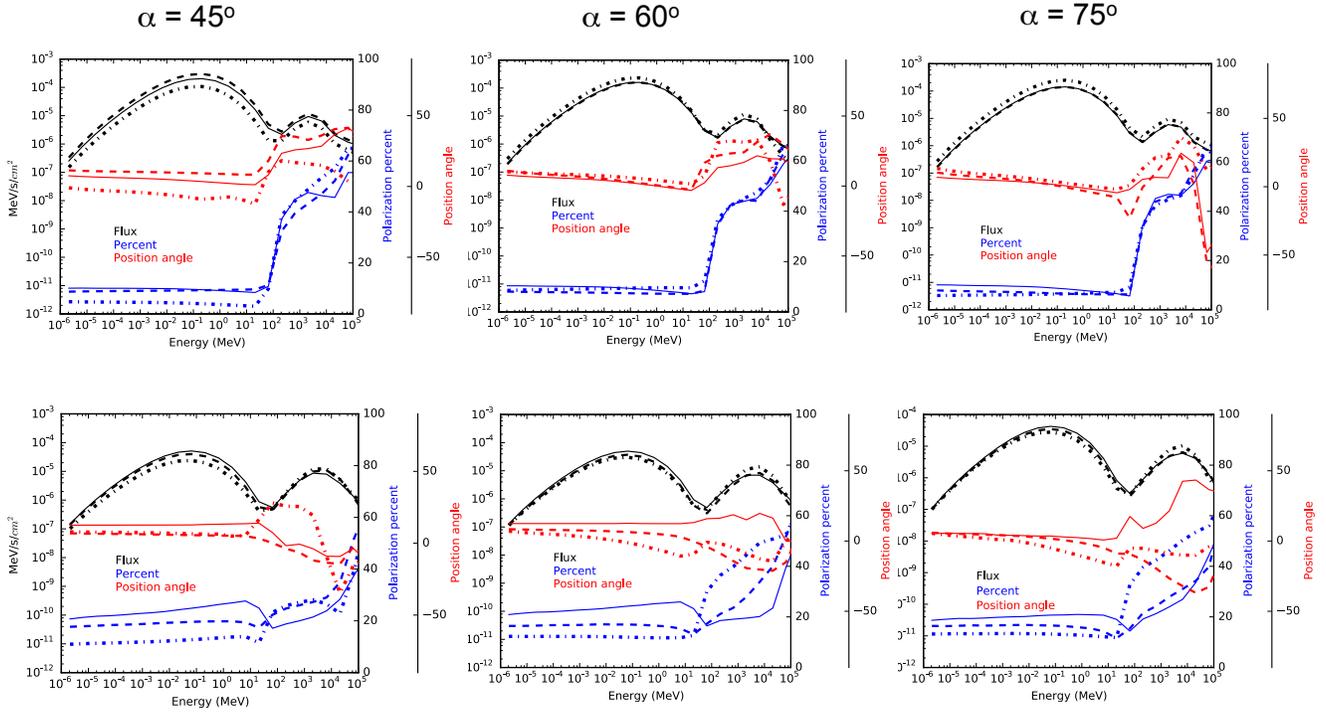}
\caption{Phase-averaged flux (black), position angle (red) and polarization percent (blue) as a function of energy for different inclination angles, as labeled, and viewing angles, $\zeta = 40^\circ$ (dot-dashed lines), $60^\circ$ (dashed lines) and $70^\circ$ (solid lines) for SR pitch angle, $\psi = 0.01$.  The top panels are for minimum emission radius $r  = 0.7 - 1.3\,R_{\rm LC}$ and the bottom panels are for $r = 1.3 - 2.0\,R_{\rm LC}$.  Flux units are the same for all panels.}
\label{fig:pa}
\end{figure}

The phase-averaged flux, polarization degree and position angle as a function of energy are calculated from Eqs \ref{eqn:Fw}, \ref{eqn:Pw} and \ref{eqn:PAw} respectively.  Figure \ref{fig:pa} shows the combined phase-averaged SR and CR spectral energy distributions (SEDs), PA and polarization percent for three different inclination angles, $\alpha = 45^\circ, 60^\circ$ and $75^\circ$, three different viewing angles, $\zeta = 40^\circ, 60^\circ$ and $70^\circ$ and two emission radius ranges, $r = 0.7 - 1.3\,R_{\rm LC}$ (top panels) and $r = 1.3 - 2.0\,R_{\rm LC}$ (bottom panels).  The SR peak in the SED at lower energy is broader than the CR peak at higher energy since it is emission from a broad range of pair energies (see HK15) while the CR is from primary particles.  The ranges of angles are chosen primarily to model the Crab pulsar, whose viewing angle has been constrained to $63^\circ \pm  1.3^\circ$ by modeling the Chandra image of the pulsar wind nebula (Ng \& Romani 2008).  The inclination angle is unconstrained by flux observations but as we will discuss, could potentially be constrained by multiwavelength polarization measurement.  

For the smaller emission radii, there is little variation with inclination angle.  The general trend is for the polarization to be quite low, $< 10$\%, at energies below the SR to CR transition around 100 MeV and to rise sharply at higher energy to about 45\% and then increasing to 60\% beyond the CR SED peak.  As discussed earlier, the low polarization degree for SR is caused both by depolarization in the caustic peaks and by the combined emission from a range of different gyro phases that have different electric vector directions.  The CR maintains the intrinsic degree of polarization outside the peaks since the acceleration and electric vector always points in the direction of the particle trajectory curvature.  The PA for SR is nearly aligned with the projection of the spin axis on the plane of the sky ($0^\circ$) but at the transition to CR the PA increases suddenly for the lower altitude emission range.  There is a decrease in PA with increasing energy, most strongly for  $\alpha = 75^\circ$.  There is little variation with viewing angle except at  $\alpha = 45^\circ$, where the PA and degree of SR polarization are smaller for $\zeta = 40^\circ$.  

\begin{figure} 
\hskip -1.5cm
\includegraphics[width=200mm]{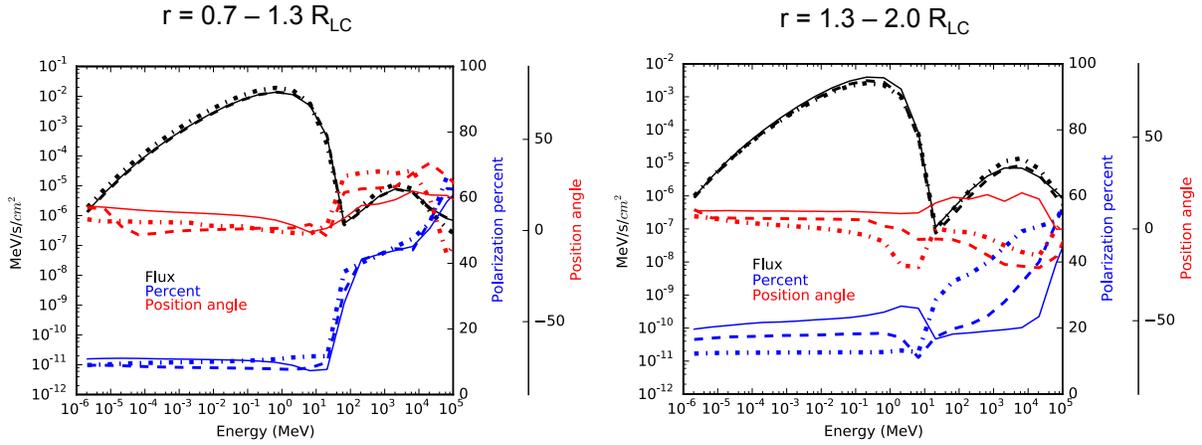}
\caption{Phase-averaged flux (black), position angle (red) and polarization percent (blue) as a function of energy for inclination angle $\alpha = 60^\circ$ and viewing angles, $\zeta = 40^\circ$ (dot-dashed lines), $60^\circ$ (dashed lines) and $70^\circ$ (solid lines) for SR pitch angle, $\psi = 0.1$.  The left panel is for minimum emission radius $r  = 0.7 - 1.3\,R_{\rm LC}$ and the right  panel are for $r = 1.3 - 2.0\,R_{\rm LC}$.}
\label{fig:papsi0.1}
\end{figure}

For the larger emission radii outside the LC, the PA for SR is again nearly aligned with the rotation axis projection, but is constant with energy except for $\zeta = 40^\circ$ at $\alpha = 60^\circ$ and $75^\circ$ where there is a decrease with increasing energy.  Above the SR-CR transition, there are significant variations with inclination and viewing angle with the PA showing 
both increases and decreases.  The SR polarization percent is higher than for the smaller emission radii, averaging around 18\% and varying significantly with viewing angle, with the lowest values of around 10\% for $\zeta = 40^\circ$ and higher values around 20\% for $\zeta = 70^\circ$.  In contrast, the CR polarization degree is lower than for the smaller emission radii, except for $\zeta = 40^\circ$ at $\alpha = 60^\circ$ and $75^\circ$.  Figure \ref{fig:papsi0.1} shows the phase-averaged polarization for pitch angle $\psi = 0.1$, $\alpha = 60^\circ$ and the two ranges of emission radii, illustrating that there is very small variation with pitch angle.

\begin{figure} 
\hskip 0.0cm
\includegraphics[width=170mm]{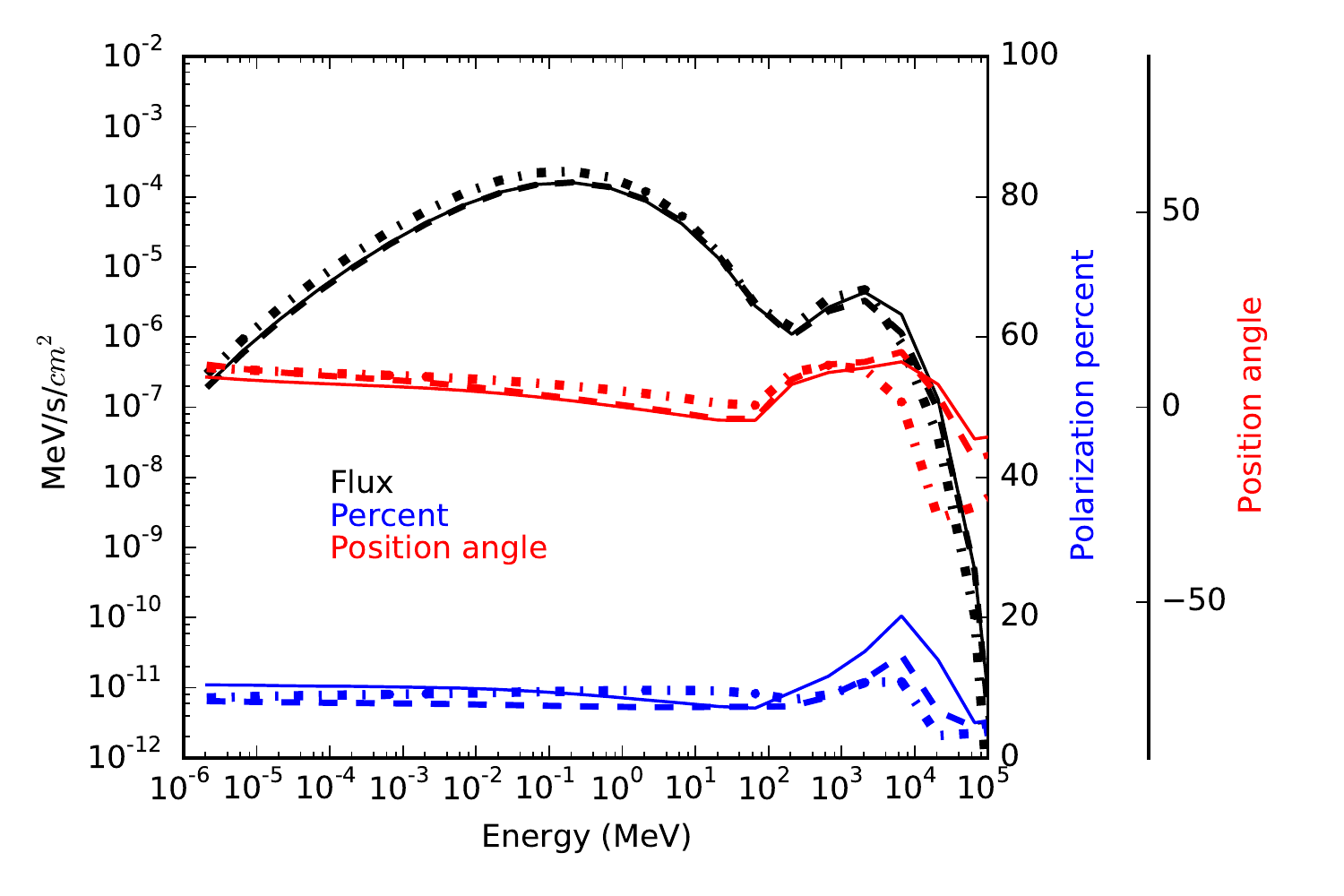}
\caption{Phase-averaged flux (black), position angle (red) and polarization percent (blue) as a function of energy for inclination angle $\alpha = 60^\circ$, and viewing angles, $\zeta = 40^\circ$ (dot-dashed lines), $60^\circ$ (dashed lines) and $70^\circ$ (solid lines), for the case where the high-energy SED component is synchrotron radiation for emission radius $r  = 0.7 - 1.3\,R_{\rm LC}$ and SR pitch angle, $\psi = 0.01$.}
\label{fig:paSR}
\end{figure}

There are generally larger changes in PA with energy for SR emission starting inside the LC where the magnetic field is poloidal and the magnetic field and electric vector directions are changing with distance from the neutron star.  For all $\alpha$, the PA of SR decreases with energy for most $\zeta$ until the jump around 10-100 MeV after which the CR PA increases with energy.   Outside the LC, the toroidal field direction is roughly constant with distance from the neutron star so that there is less change of PA with energy for both SR and CR.  
It would thus be possible to constrain the emission radius and whether the emission mechanism at $\gamma$-ray energies is SR or CR by measuring the polarization at different energies.  
Detecting a change is PA as a function of energy below 10 MeV would indicate that the pair SR originates at least partly inside the LC.   Detecting a large increase in polarization degree between hard X-ray and $\gamma$-ray bands and a jump in change PA 
would suggest that the $\gamma$-ray emission is CR.  To test this hypothesis, we have simulated the phase-averaged broadband polarization for the case where both pairs and primary particles emit SR by artificially suppressing primary CR and giving the primary particles pitch angles of $p_\perp = 5 \times 10^{-5}\gamma$ for emission radii  $r = 0.7 - 1.3\,R_{\rm LC}$.  The result, shown in Figure \ref{fig:paSR}, is that the PA changes by only a small amount over the entire energy range, with the exception of a small dip at the transition between the two SED components and the polarization degree remains low at all energies.   Increasing the radial extent of the emission increases the effects of depolarization of SR and the jumps of polarization degree at the SR/CR transition.  For minimum emission radii below $0.5\,R_{\rm LC}$, the phase-average polarization degree drops below 5\%, which could not agree as well with the pulsar observations discussed below.
 
The phase-averaged polarization of the Crab pulsar and nebula has been measured in optical and X-ray bands and the results depend on field-of-view (FOV).  
Measurements with HST (Moran et al. 2013a,b) for a FOV of $1.7' \times 1.7'$ centered on the pulsar gives a PA $= 155^\circ$ and P = 19\%, and for a smaller FOV of $2'' \times 2''$ that includes the pulsar and inner knot gives PA $= 105^\circ$ and P = 5.2\%.  Measurement of the inner nebula by INTEGRAL (Dean et al. 2008) for a FOV of $25'' \times 33''$, that includes the pulsar, inner knot, X-ray jet and wisps, gives PA $= 123^\circ\pm 11^\circ$ and P = $46 \pm 10$\%.  The PA of the inner nebula is consistent with the projected direction of the pulsar rotation axis of $124^\circ$, while the PA $= 105^\circ$ of the pulsar may be aligned with the pulsar proper motion vector, PA $= 110 \pm 2 \pm 9^\circ$ (where both measurement and reference frame uncertainties are given) (Kaplan et al. 2008).  
Slowikowska et al. (2009) obtain a phase-averaged polarization percent of only 5.5\% after subtracting an assumed constant nebular component of 33\%.  Forot et al. (2008) obtain a much higher phase-averaged pulsar polarization fraction of $0.47^{+0.19}_{-0.13}$ at soft $\gamma$-ray energies with INTEGRAL/IBIS (200 keV - 5 MeV), where the bridge and off-peak flux of the pulsar is much higher. 
To compare with our results, the measured sky projection of the pulsar spin axis of $124^\circ$ corresponds to a PA of $0^\circ$ in our models.  Our results for the SR phase-averaged PA in the optical band is a few degrees larger than $0^\circ$ which is roughly consistent with observations.  The measured low degree of polarization of the pulsar emission at optical energies is more consistent with our result of $< 10$\% polarization for the lower range of emission radii starting inside the LC in our model.  Our result of $\sim 20$\% polarization for the emission range $r = 1.3 - 2.0\,R_{\rm LC}$ indicates that the pulsed emission does not come from only far outside the LC.  Measuring a trend in PA with energy above the optical band could further discriminate between different inclination angles and emission radii.  

\section{Discussion}   \label{sec:diss}

We have presented the broadband energy-dependent polarization characteristics of a rotation-powered pulsar that emits a combination of SR at optical to hard X-ray energies and CR or SR at GeV $\gamma$-ray energies.  Such a combination of emission mechanisms may be characteristic of both Crab-like pulsars and a number of millisecond pulsars.  Our results for emission starting both inside and well outside the light cylinder in a force-free magnetic field geometry give different polarization characteristics in both phase-resolved and phase-averaged emission that could be used to constrain the location of the emission, the system geometry and the emission mechanisms.  The phase-resolved polarization of the Crab pulsar has been well measured at optical wavelengths (Slowikowska et a. 2009) and displays deep depolarization dips as well as sharp swings of PA at the light curve peaks.  As we and others (Dyks et al. 2004, Petri 2011, Cerutti et al. 2016), have shown, these are expected characteristics of caustic emission from the pulsar outer magnetosphere and current sheet.  Those papers however modeled either mono-energetic radiation or only the $\gamma$-ray band.  Our multiwavelength treatment provides the prospect of using measurements in different energy bands to probe characteristics of the pulsar emission.  Phase-resolved and phase-averaged polarization measurements of the pulsed radiation in the X-ray band, together with the optical measurements, can constrain the inclination and viewing angles as well as the location of the emission (inside or outside the LC).  In the case of the Crab, where the viewing angle is already constrained by the X-ray torus modeling, the inclination angle can be even better constrained.  
Measurement of the polarization in the 1 - 100 MeV band could distinguish whether CR, SR or inverse Compton is responsible for the $\gamma$-ray emission.  If a sudden change in PA and a sharp rise in polarization degree between different spectral components is observed, then CR would be strongly indicated as the $\gamma$-ray emission mechanism since SR is most likely responsible for the X-ray emission.  Outer gap (Takata \& Chang 2007) and slot gap (Harding et al. 2008) models for emission inside the LC, as well as models for emission outside the LC near the current sheet (Kalapotharakos et al. 2014, HK15, Kalapotharakos et al. 2017) predict that CR is responsible for GeV $\gamma$-ray emission.  Another group of models for emission from the striped wind (Petri \& Kirk 2005) or reconnection in the current sheet  outside the LC (Uzdensky \& Spitkovsky 2014, Cerutti et al. 2016a,b) predict that SR is the GeV emission mechanism.  Lyutikov (2013) predicts that the GeV emission is inverse Compton emission,  and in the outer gap model of Takata \& Chang (2007) the emission between 10 and 100 MeV is inverse Compton and is found to have very low ($< 10$ \%) polarization degree.
A polarization measurement in the $\gamma$-ray band would rule out some of these models since only CR is able to produce high degrees of polarization above 20\% and as high as 40\%.  Our results predict that a decrease in phase-averaged PA with energy from optical through hard X-rays should be observed for emission starting inside the LC while little or no energy dependence of their PA is predicted for emission outside the LC.   Although comparison of measured PA by different detectors in separate energy bands could constrain the emission location, ideally such a measurement is best made by a single detector with a broad energy sensitivity so that different FOV is not an issue.  

A number of existing and planned X-ray polarimeters offer improved sensitivity in soft X-ray to $\gamma$-ray bands.  X-Calibur is a balloon-borne focussing hard X-ray telescope that can  measure polarization in the 20 - 80 keV band (Beilicke et al. 2014).  It had a recent launch in September 2016 and is expected to have several additional launches over the next few years.  The CZTI instrument on Astrosat, launched in 2015, will measure polarization in the 100 - 380 keV band (Vadawale et al. 2015).
In the soft X-ray band (2 - 10 keV), {\sl IXPE} (Imaging X-ray Polarimetry Explorer) (Weisskopf et al. 2013) has been selected  by NASA as a SMEX mission and will be able to reach a Minimum Detectable Polarization (MDP) of 5\% for bright sources.  At higher energies, AMEGO (All-Sky Medium Energy Gamma-Ray Observatory)\footnote{https://asd.gsfc.nasa.gov/amego/index.html}\footnote{https://pcos.gsfc.nasa.gov/physpag/probe/AMEGO\_probe.pdf} is a proposed NASA probe mission that will have sensitivity to polarization in both the Compton scattering (0.2 - 10 MeV) and pair production (10 MeV - 10 GeV) regimes, using silicon strip trackers and CZT and CSI calorimeters.  This band that is particularly important for observing the transition between X-ray and $\gamma$-ray emission mechanisms.  
The gas Time Projection Chamber (TPC) technique has also been proposed for high-energy pair production polarimeters.  AdEPT is a proposed TPC $\gamma$-ray polarimeter with a projected MDP of 5\% at 1 - 100 MeV (Hunter et al. 2014), HARPO is another proposed TPC polarimeter that would be sensitive in the MeV - GeV energy range (Denis et al. 2014).
It is also possible that the {\sl Fermi} Large Area Telescope, although not designed for polarimetry, will be sensitive to high degrees of polarization (30\% - 50\%) above 200 MeV after 10 years of observations of bright pulsars such as Crab and Vela (Giomi et al. 2016).  Such a detection would strongly suggest that CR is at least partly responsible for the observed $\gamma$-ray emission.  The multiwavelength polarization predictions we have presented for a generic emission model should be testable with data from these instruments.

\acknowledgments  
This work is supported by the National Science Foundation under Grant No. AST-1616632 and by the NASA Astrophysics Theory Program.  Resources supporting this work were provided by the NASA High-End Computing (HEC) Program through the NASA Center for Climate Simulation (NCCS) at Goddard Space Flight Center.   

\newpage

\newpage

\newpage

\newpage

\newpage

\newpage

\newpage


\end{document}